\documentclass[11pt]{spie} 
\usepackage{subfig}
\usepackage{amsfonts}
\usepackage{amsmath}
\usepackage{amssymb}
\usepackage{calc}
\usepackage{authblk}
\usepackage{graphicx}
\newcommand{\vi}{``}
 \newcounter{thm}
 \newcounter{ex}
 \newcounter{re}
\title{The role of complex analysis in modeling economic growth}
\author{Angelica Sbardella$^{1,2}$, Emanuele Pugliese$^{3}$, Andrea Zaccaria$^{2}$, and Pasquale Scaramozzino$^{1,4}$*}

\affil[$^{1}$]{ Department of Economics and Finance, Universit\`a degli Studi di Roma Tor Vergata, Rome, Italy}
\affil[$^{2}$]{ ISC-CNR -- Institute of Complex Systems, Rome, Italy}
\affil[$^{3}$]{ JRC, European Commission, Seville, Spain}
\affil[$^{4}$]{ School of Finance and Management, SOAS University of London, London, UK\\ *ps6@soas.ac.uk}

\begin{document}
\maketitle
\abstract{Development and growth are complex and tumultuous processes. Modern economic growth theories identify some key determinants of economic growth. However, the relative importance of the determinants remains unknown, and additional variables may help clarify the directions and dimensions of the interactions. 
The novel stream of literature on economic complexity goes beyond aggregate measures of productive inputs, and considers instead a more granular and structural view of the productive possibilities of countries, \textit{i.e.} their capabilities. Different endowments of capabilities are crucial ingredients in explaining differences in economic performances.  In this paper we employ economic fitness, a measure of  productive capabilities obtained through complex network techniques.  
Focusing on the combined roles of fitness and some more traditional drivers of growth, we build a bridge between economic growth theories and the economic complexity literature.
Our findings, in agreement with other recent empirical studies, show that fitness plays a crucial role in fostering economic growth and, when it is included in the analysis,  can be either complementary to traditional drivers of growth or can completely overshadow them. 
}

\keywords{Economic Fitness; Complexity; Capabilities; Economic Growth.}

\section{Introduction} \label{sec:intro}
Why are some countries wealthier than others? And why do some countries exhibit sustained rates of growth over long periods, whereas others appear to be stuck in a low-income, low-growth path? These questions have been central to economics ever since its origin as a science, following Adam Smith’s [\citen{smith}] original enquiry. An understanding of the main determinants of long-run growth arguably remains the most important issue in economics.

The ultimate causes of economic growth are however still not fully understood. In an influential study, Helpman [\citen{helpman2005mystery}] examines the recent literature on the subject and acknowledges that the determinants of economic growth remain a mystery. Early models based on capital deepening and exogenous technical progress have given way to contributions that emphasise the endogenous nature of economic growth, and that explore the role of factors such as expenditure on education, investment in research and development, openness to international trade, and the presence of institutions that foster social and economic inclusion [\citen{aghion98growth}]. 

Whilst all of these factors play a role in affecting economic growth, the relative importance of each of them is more difficult to establish, especially because they are bound to interact with each other in complex ways. Furthermore, this list of key determinants of economic growth may not be exhaustive: additional factors which have hitherto been ignored may be just as important, and may contribute significant predictive power to the models that have been studied so far.
Exogenous as well as endogenous growth theories predict that growth can be explained by a set of variables which capture both the initial conditions of the economy and the rate at which its production inputs are accumulated. The extent to which these models can predict the growth performance of broad cross-sections of countries or regions is however limited [\citen{barro2004economic,acemoglu2009introduction}]. It is therefore important to look for additional drivers of growth, which may have been overlooked in the more traditional analyses.

The present paper explores one such additional factor: the fitness of the economy, as measured by a complex-network metric based on the countries’ revealed comparative advantage [\citen{tacchella12}]. The profile of trade specialization of a country can in fact be regarded as a reflection of its underlying capabilities, which can be defined as the skills that enable its economy to expand into new production requirements and to adopt new technologies [\citen{teece1997dynamic}]. Conventional measures of the production possibilities of an economy do not explicitly consider its degree of flexibility and adaptability: if more complex economies are however also endowed with a richer set of capabilities, then ignoring them would lead to a misspecification of the underlying models, because relevant explanatory variables would be omitted from the analysis.

The structure of this paper is as follows. Section \ref{sec:ecgr} discusses the role of capabilities and complexity for economic growth. Section \ref{sec:methods} defines the measure of Fitness used in the paper and explains the empirical methodology and the data. Section \ref{sec:evid} presents the empirical evidence from both non-parametric graphical analysis and econometric estimation. Section \ref{sec:concl} concludes.

\section{Economic Growth, Capabilities and Complexity}\label{sec:ecgr}
In his encyclopaedic treatment of modern theories of economic growth, Acemoglu [\citen{acemoglu2009introduction}] offers a thorough overview of the main approaches that have been set out to explain the dynamic growth paths of the economies in the long run. The early analysis by Solow [\citen{solow56}] identified the main sources of growth in the accumulation of physical capital and in exogenous technical progress, which is responsible for the upward shift of the production possibility frontier of the economy over time. In the presence of decreasing returns to capital in the aggregate production function, Solow’s model predicts that the rate of growth will eventually tend to peter out, as the economy approaches its long-run steady-state dynamic equilibrium path.

An important implication of the exogenous growth model is that the income per capita of low-income economies should tend to converge to that of high-income ones in the long run, and we should therefore observe a catching-up of less developed economies to the more developed ones. Empirical analyses show that there is some evidence of convergence of low income to high income economies, albeit at a very slow pace [\citen{barro2004economic}].

More recent analyses of growth have sought to endogenize the rate of technical progress as the outcome of explicit decisions made by economic agents. These models overcome the assumption of decreasing returns with respect to the factor that is accumulated by acknowledging to the existence of externalities among firms. This view was first explored by Arrow  [\citen{arrow1962economic}], and is captured in his notion of learning-by-doing. New technologies are incorporated in new investments, and therefore the rate at which aggregate productivity increases is directly related to the rate of new investment in the economy. As a result, the productivity of individual firms is an increasing function of the aggregate capital stock in the economy. This way, it is possible to have constant or even increasing returns to capital accumulation at the aggregate level, even if the technology at the firm level still exhibits conventional decreasing returns to its own production inputs.

Later developments in endogenous growth theories identify the main drivers of long-run growth in investment in education in the presence of externalities from human capital accumulation  [\citen{lucas1988mechanics}], in expenditure in Research and Development in a stochastic Schumpeterian model of creative destruction [\citen{aghion1992model}], or in the openness of economy to international trade through learning-by-exporting [\citen{frankel1999does,bernard2003plants}].  These drivers may interact with each other, as in the model by Lucas  [\citen{lucas1993making}] where a high level of human capital in the economy facilitates investment in new technologies, and in turn enhances their growth-improving effects.
The recent literature on growth has also explored the role of social and political institutions. In particular, Acemoglu and Robinson [\citen{acemoglu12}] distinguish between inclusive institutions, which stimulate entrepreneurship and innovations, and extractive institutions, which are instead responsible for creating incentives to exploit rents and which have a dampening effect on growth.

Endogenous growth theories are able to explain the lack of convergence of poorer countries and regions to the richer ones: since the returns to the factors which are accumulated are now constant or even increasing, the rate of growth of the economy is not necessarily predicted to slow down along its long-run equilibrium path.
An important extension of growth models further considers the possibility of multiple steady state equilibria. Murphy, Shleifer and Vishny [\citen{murphy1988}] formalise the notion of a big push, according to which a discrete development effort is required for a low-income economy to escape its poverty trap and to set in motion a process of self-sustaining growth. The rationale for multiple equilibria rests on the assumptions of increasing returns in the scale of production and of non-pecuniary externalities among sectors. Multiple equilibria can provide a justification for targeted development policies, for instance in the form of government support to industrialisation in a critical mass of industrial sectors. Models with multiple steady-state equilibria have been generalised to models with multiple dynamic steady states (e.g. Krugman, [\citen{krugman1991history}], and Matsuyama, [\citen{matsuyama1991increasing}]) where initial conditions can be critical to determine the long-run growth path of the 
economy.

A novel approach to the analysis of economic growth, however, goes beyond aggregate measures of the production inputs and considers instead a more granular and structural view of the production possibilities of the economy. This approach examines the possible role of capabilities, which could be defined as a broad set of skills that could adapt to changing production requirements and which facilitate the introduction of new technologies. This approach can be traced back to the seminal work by Hirschman [\citen{hirschman1958strategy}], where capabilities make it possible to create backward and forward linkages across economic sectors, and Penrose [\citen{penrose1959theory}]  with her resource-based theory of the firm. Teece \textit{et al.} [\citen{teece1994coherence}] argue that capabilities in an organization are both intangible and non-transferable, and lay the foundations for the modern complex network analysis by showing that they constitute the key factors for a coherent growth of the firm 
consistent with its core competencies. Abramovitz [\citen{abramovitz1986catching}] referred to social capabilities as all those attributes that affect a country’s ability to operate modern and large-scale
businesses, and which include their political and social characteristics: they should therefore not be interpreted in a narrow individualistic sense (see also Fagerberg and Srholec, [\citen{fagerberg2017capabilities}]). Capabilities can affect economic growth directly through their impact on the adoption of innovations, as shown by Lall [\citen{lall1992technological}] and by Kremer [\citen{kremer1993ring}].

In a series of important contributions, Sutton [\citen{sutton2002rich,sutton2012competing}], Sutton and Trefler, [\citen{sutton2016capabilities}]) argues that the source of the main differences in output per capita across economies lies not in their accumulation of physical factors of production, as in the conventional theories of growth, but rather in the set of capabilities with which their economy is endowed. These capabilities enable firms to take advantage of investment by increasing their labour productivity and also make it possible to further expand their set of skills, thereby generating a virtuous circle. Sutton’s analysis combines insights from the Industrial Organization literature, the Economic Geography literature, and the Trade literature. Central to his results is that, in equilibrium, the market structure of an industry will tend to encompass a variety of products of different quality. The capabilities required to produce high-quality products will always be scarce. This implies that, on 
a global scale, in each industry there will only be a limited number of firms which hold a dominant position. From Economic Geography, capabilities must be concentrated in some countries in order to take full advantage of agglomeration economies. Each country’s comparative advantage will therefore reflect what Sutton calls the \vi economics of quality''. The available capabilities become a key factor in global trade specialization and in determining the growth prospects of each country.

Capabilities are however unobservable. Hausmann, Hwang and Rodrik [\citen{hausmann2007you}] argue that they can be inferred from a country’s export specialization profile. Production of complex goods involves the joint execution of a large number of highly specialised tasks [\citen{kremer1993ring,dalmazzo2007ring}]. This requires both the existence of broad sets of advanced skills, and the ability to combine them effectively. A detailed examination of the revealed comparative advantage across all countries in the world using the tools of complex network analysis can identify those countries which specialise in complex products. Their production 
could involve a number of tasks by high-skilled workers, who could not be substituted by low-skill workers without compromising the quality of the finished product [\citen{kremer1993ring}]. The sophisticated production structure of these countries also allows them more easily to introduce innovations and to adopt advanced technologies consistent with the role of dynamic capabilities in an environment of rapid technological change analyzed by Teece \textit{et. al. }[\citen{teece1997dynamic}]. The location of a country in the global product space would therefore be a significant predictor of its growth potential.

The usefulness of measures of product complexity for predicting growth has been demonstrated by Hausmann, Hwang and Rodrik [\citen{hausmann2007you}], who show that the mix of good that a country produces can explain its growth rate. Ferrarini and Scaramozzino [\citen{ferrarini2016production}] confirm that production complexity can help explain differences in economic performance, in an endogenous growth model with human capital accumulation. Pugliese\textit{ et al.} [\citen{pugliese2017complex}] analyze the role of fitness during the process of development and industrialization showing that fitness crucially eases the passage towards sustained growth.
 Sbardella  \textit{et al.} [\citen{sbardella2017economic}] empirically demonstrate that fitness is a key variable in explaining the relationship between wage inequality and economic development. 
Boltho \textit{et al.} [\citen{boltho2018east}] argue that the more complex production structure in East Germany relative to that in the Italian Mezzogiorno can contribute to explain the unsatisfactory performance of the latter, and its lack of convergence to the more prosperous regions in the Centre-North of the country. 

Cristelli \textit{et al.} [\citen{cristelli2017predictability}] develop an innovative approach to predict economic growth, the Selective Predictability Scheme (SPS), which makes use of ideas from the theory of economic complexity and which adopts the the fitness metric [\citen{tacchella12,tacchella13,cristelli13}]. SPS is shown to out-perform a number of alternative forecasting models, thus confirming the key role played by the complexity of the structure of production for predicting the future economic performance of a country.  Tacchella  \textit{et al.} [\citen{tacchella2018dynamical}] further develop this approach to growth forecasting and outperform the accuracy of the IMF five-year forecasts by more than $25\%$ .
\section{Materials and Methods}\label{sec:methods}
\subsection{Measuring Fitness}
As mentioned above, in this paper we use the fitness [\citen{tacchella12,tacchella13,cristelli13}], a measure of economic complexity based on cross-country differences in productive structures. Fitness is a proxy for capabilities, and is calculated by applying complex network techniques to the analysis of international trade data. 
In fact, GDP may not be sufficient to describe development and growth processes, as two countries with similar GDP levels may actually possess profoundly different endowments of capabilities. Countries that are below the income expected from their economic performances may have  already developed the full range of products that is within their technological reach, nevertheless this capability level may have not been yet translated into higher GDP levels. Thus, as explained in Section \ref{sec:ecgr}, different endowments of capabilities are the main, albeit not empirically observable, sources in explaining different economic performances and in shaping the export profiles of countries.

\begin{figure}[h]
\centering 
\includegraphics[width=.75\linewidth]{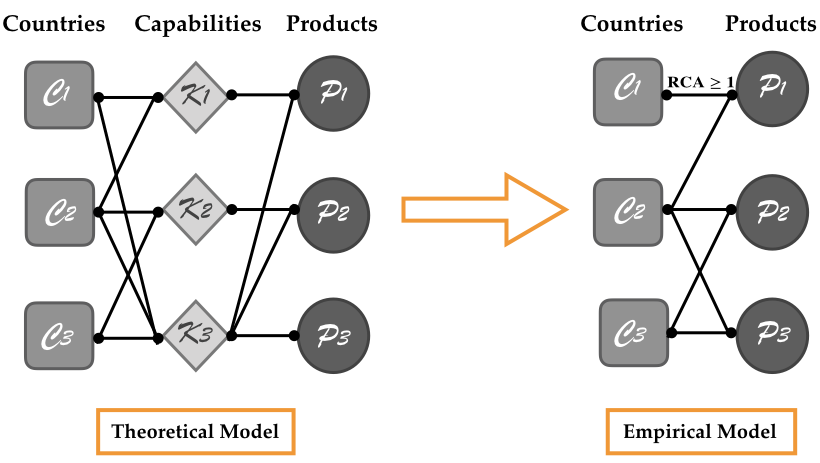}
\caption{\small In our theoretical framework capabilities are crucial in explaining the economic performances of countries. As can be observed on the left,  conceptually we can visualize capabilities as an intermediate level in a tripartite country-capability-product network. Nevertheless, this is a theoretical model,  capabilities in fact are non-measurable entities. Information on capabilities can be gathered by building an empirical country-product network through international trade data. As can be seen on the left, such bipartite network can be interpreted as the projection of the tripartite network. In the country-product network a country-product link is established if and only if the  country has a revealed comparative advantage in exporting that product.  [\citen{tacchella13,cristelli13}].   \label{fig:trip-netw}}
 \end{figure}

In network theory, the notion of capabilities can be conceptually described as an intermediate network layer that connects countries to their exported products [\citen{tacchella13,cristelli13}]. As schematically illustrated in Figure \ref{fig:trip-netw}, we can define a tripartite network whose three classes of nodes are countries ($\mathcal{ C}$), capabilities ($\mathcal{K}$) and products ($\mathcal P$). The allowed links, connecting the $\mathcal{K}$ nodes to the $\mathcal{C}$ or $\mathcal{P}$ nodes, describe the capability owned by a country and the capability required to export a product with a comparative advantage. Tacchella \textit{et al.} [\citen{tacchella13}] explore the functioning of this tripartite network and show the high degree of correlation between a country's endowment of capabilities and its fitness. 
However, the capability layer, despite being conceptually crucial, is intangible and the tripartite network is a purely theoretical tool to better visualize our theoretical framework. By using  international trade data,  the information coded in the hidden capability layer can be gathered by building an empirical bipartite network in which countries are connected to the products they export [\citen{hidalgo09}] with a revealed comparative advantage (RCA) [\citen{balassa}].  As represented in Figure \ref{fig:trip-netw},  this country-product network is viewed as a contraction of the tripartite network over the capability dimension.
How to extract in an optimal way the informative content coded in such country-product network?
The fitness-complexity metric provides the correct mathematical specification based on the network topology, and, through an iterative algorithm, it defines a measure of country competitiveness, \textit{i.e.} fitness, and of product sophistication, \textit{i.e.} complexity. 
To understand the rationale beyond the metric it is useful to observe the binary adjacency matrix of the network, $\hat M$, whose rows represent countries and columns their exported products (see Fig.\ref{fig:mcp-fq-mr}). RCA is used to filter and digitize the data allowing to focus on qualitative differences, rather than quantitative, in the export baskets of countries: which kinds of products are exported competitively, not in which volumes. Therefore, matrix entries are set equal to 1 when a country exports a product with $RCA \geq 1$, and 0 in the opposite case:

\begin{eqnarray}
M_{cp}=
\begin{cases}
1 &\text{ if $RCA_{cp} \geq 1$}\\ 
0  &\text{ if $RCA_{cp} < 1$.}\\
\end{cases}
 \end{eqnarray}
 
\begin{figure}[h]
\centering 
\includegraphics[width=.75\linewidth]{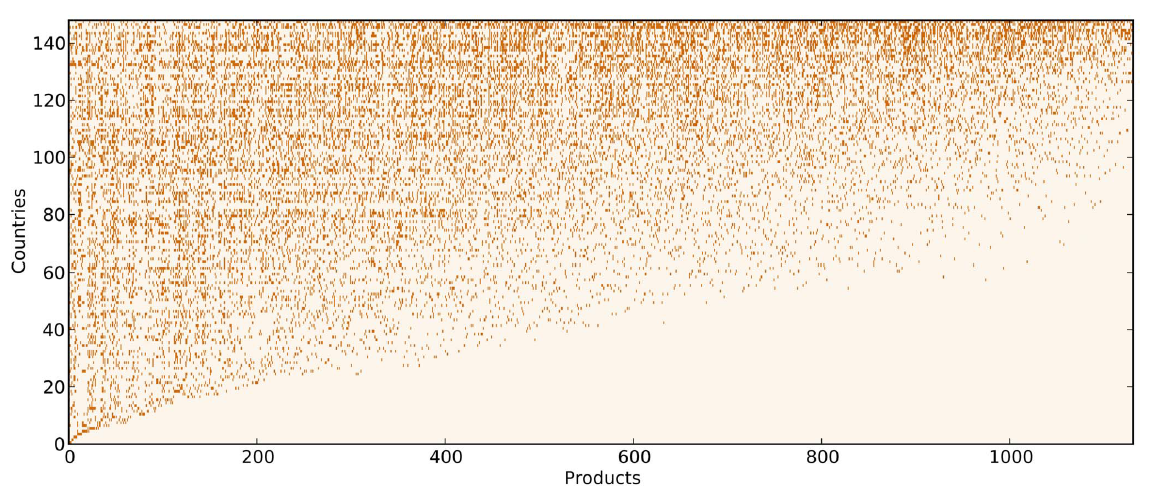}
\caption{\small The binary matrix of countries and products built from the worldwide 1998's export flows of BACI data-set [\citen{baci}].
The rows and columns of the matrix are ranked according to the fitness-Complexity algorithm. The rows are sorted by increasing country fitness and the columns by increasing product Complexity. In such a way, the matrix acquires a triangular-like shape: countries with more diversified export baskets are more competitive, while countries specialized in few products --which generally are also exported by every other country-- are the less competitive. Source of the figure: [\citen{tacchella13}]. }
 \end{figure}\label{fig:mcp-fq-mr}
If suitably ordered, the matrix assumes a quasi-triangular shape. 
Indeed, by looking at the matrix triangularity, it is possible to infer that, on the one hand, a largely diversified country's ability to export a product with a comparative advantage gives no clues on its complexity. On the other hand, when a poorly diversified country --which has a comparative advantage in exporting few and very ubiquitous products-- is able to export a product with a comparative advantage it is likely that its production requires a low level of industrial sophistication. This is quite informative: a product is complex if low-fitness countries do not export it. 
To make this clearer, consider a straightforward example of a high complexity product, transistors, and a low complexity product, nails. Only highly industrially and technologically developed countries are able to export transistors; by contrast, nails are exported by all sorts of countries, both more and less industrialized. Consequently, the low complexity of nails can be surmised directly from their presence in the export basket of low fitness countries. 
This observation hints at a non-linearity in the relation between product Complexity and country fitness\footnote{The need for a non-linear coupling between fitness and Complexity was formalized in mathematical terms by Caldarelli et al. [\citen{caldarelli2012network}].}. Therefore, from the matrix $\hat M$, it is possible to obtain an intensive metric that measures country fitness ($F_c$) as a weighted average of the country export basket's diversification, where the weight is the complexity associated to each product. 
Product Complexity (${Q_p}$), instead, is calculated as the number of countries that export the product with comparative advantage, bounded by the fitness of the least competitive exporter of the product. In formula:

\begin{eqnarray}\label{pilrs1}
\begin{cases}
\widetilde{F}_c^{(n)}=\sum_p M_{cp} Q_p^{(n-1)} & \\ \\
\widetilde{Q}_p^{(n)}=\dfrac{1}{\sum_c M_{cp} \dfrac{1}{F_c^{(n)}}} \\
\end{cases}
\begin{cases}
F_c^{(n)}=\dfrac{\widetilde{F}_c^{(n)}}{<\widetilde{F}_c^{(n)}>_c}  & \\ \\
Q_p^{(n)}=\dfrac{\widetilde{Q}_p^{(n)}}{<\widetilde{Q}_p^{(n)}>_p}\\ 
\end{cases}\label{eq:f-q}
\end{eqnarray}
where $<\cdot>_x$ denotes the arithmetic mean with respect to the possible values assumed by the variable dependent on $x$, with initial condition:
\begin{equation}\label{pilrs2}
\sum_p Q_p^{(0)}=1   \hspace{6pt}\forall \, p.
\end{equation}

The iteration of the coupled equations leads to a fixed point which has been proved to be stable and non-dependent on initial conditions [\citen{tacchella12}]. The fixed point defines the non-monetary metric which actually quantifies $F_c$ and $Q_p$. The convergence properties of Eq.\ref{eq:f-q} are not trivial and have been extensively studied by Pugliese et al. [\citen{pugliese14conv}].
\subsection{Econometric Model and Graphical Representation}

As we have discussed in the previous sections, the economic complexity approach  has a structural interpretation of growth and development, understood as the outcomes of a learning process through which new capabilities are added to the existing pool thus opening up new and more complex productive possibilities which will eventually lead to higher prosperity and faster economic growth.

In this section, to partially reconcile this new view on development with the stylized facts of neoclassical theories of economic growth we integrate the economic complexity discourse with the analyses of growth determinants popularized in the 1990s by Barro [\citen{barro1991economic}], and Mankiw, Romer and Weil [\citen{mankiw1992contribution}], the so-called \vi growth regression models''.  In the latter, GDP per capita growth is decomposed into contributions associated with changes in factor inputs, production technologies, demographic variables and so on. 
We therefore focus on the multifaceted relations between GDP per capita growth, fitness, and various economic indicators considered crucial drivers for economic growth: Capital Intensity, Capital-Output Ratio, GDP per capita, Life Expectancy, Secondary Schooling, and Total Factor Productivity.
The choice of such drivers of growth is rooted in the growth regression models; in particular, the income and capital deepening variables are directly linked to the early analysis of Solow [\citen{solow56}], while the inclusion of data on educational attainment and life expectancy among the explanatory variables is based on the more recent endogenous growth theories mentioned in Section \ref{sec:ecgr}.


Following the lines of Cristelli et al [\citen{cristelli2017predictability}], we opt for a non-parametric description since the dependencies are fundamentally dynamic and non linear, and different types and shapes of relationships between fitness, the chosen growth determinants and economic growth coexist depending on the phase\footnote{Where with economic phase we refer to the definition of Cristelli et al. [\citen{cristelli15}]. These different economic regimes are explicitly visible in the fitness-GDP per capita plane.} that an economy is going through: in each phase the dimensions of interest combine in an \textit{a priori} unknown fashion.

To provide a basis for comparability with the more traditional analysis on growth determinants, we offer an additional tool of interpretation and perform a linear parametric regression model. In the latter various drivers of economic growth figure as right-hand-side explanatory variables affecting the growth rate of GDP per capita,  which is on the left-hand side of the equation.  The idea is to add fitness to the growth drivers and observe whether and to what extent the results change. As discussed above, a non-parametric approach is the most appropriate framework to investigate the complex and non-linear interactions that are at the basis of the process of growth and development. Indeed, a linear parametric approach, looking at averaged interactions and having strong assumptions on the functional forms of the relationships, might not recognize the importance of a strong but non-linear signal. Specifically, for instance, a discrete element as the presence or absence of an enabling capability necessary to 
competitively export a product can change in unpredictable ways the development profile of a country. These nuances might remain unobserved when using a linear parametric approach.
Nevertheless, since non-parametric analyses can be performed only by looking at the simultaneous effect of fitness and a single growth driver on GDP growth rate, a linear regression enables us to test the robustness of our results by  observing cross-media effects.

Our empirical evidence is obtained by using an unbalanced panel of countries over the time period 1963--2000, the number of countries slightly varying around 144.  In order to avoid cyclical short-run fluctuations, the GDP per capita growth rate is calculated over a $\Delta t=5$ years. Furthermore, to reduce the risk of simultaneity bias, a lag of
 five years is considered between the explanatory variables and GDP per capita growth rate.

In practice, in our non-parametric analysis, we explore the empirical relations by looking at tridimensional color-maps, smoothed representations of GDP per capita growth rate for different values of fitness on the x-axis and, on the y-axis, of each one of  Capital Intensity, Capital-Output Ratio, GDP per capita, Life Expectancy, Secondary Schooling, and Total Factor Productivity.  
The color-maps are realized trough a multivariate Nadaraya-Watson regression [\citen{nadaraya}], a continuous non-parametric kernel estimator  and by pooling all countries and years in the panel.

We then perform a more traditional parametric linear regression analysis. We opt for a fixed effect analysis with robust standard errors. Our explanatory variables are the one listed above, while on the left hand side of the equation figures GDP per capita growth rate. 

\subsection{Sources of data}\label{ssec:data}
The levels of GDP, Population, Total Factor Productivity, Capital, and Human Capital are taken from the Penn World Table 9.0 (PWT) produced by the \textit{University of Groningen} and the \textit{University of Pennsylvania} [\citen{pwt}]. 

We also include in our analysis the Life Expectancy 
measure of the World Development Indicators, the collection of development statistics of the World Bank [\citen{wdi2017}]. 

Finally, fitness is taken from the Export fitness data-set developed by the PIL group of the Institute of Complex Systems in Rome which covers a number of countries varying between 130 and 151, over the period 1963--2000. As briefly explained in the previous section and shown in detail in Tacchella \textit{et al.}[\citen{tacchella12}], Cristelli \textit{et al.} [\citen{cristelli13}], fitness is obtained trough an empirical algorithm that, for the chosen time window, employs international trade data from UN-COMTRADE [\citen{feenstra2005world}].
The choice of the time period of our analysis was constrained by the availability of fitness data. 

\section{Empirical Evidence}\label{sec:evid}
\subsection{Non-parametric Graphical Analysis}

\subsubsection*{Fitness and GDP per capita}

As follows from the hypothesis of diminishing returns with respect to the factor which is being accumulated, exogenous growth theories predict higher growth in response to lower starting GDP per capita, since in the long-run all economies should catch-up and converge to similar income levels. Initial GDP per capita, thus, should have a negative impact on economic growth. Such conditional convergence has been reported, even if not to the extent predicted by neoclassical models , by Barro [\citen{barro1991convergence}] and Mankiw et. al [\citen{mankiw1992contribution}]. However, other important empirical contributions do not find evidence of conditional convergence [\citen{acemoglu2009beyond, quah1996twin}]. 

\begin{figure}[h]
\centering 
\includegraphics[width=.85\linewidth]{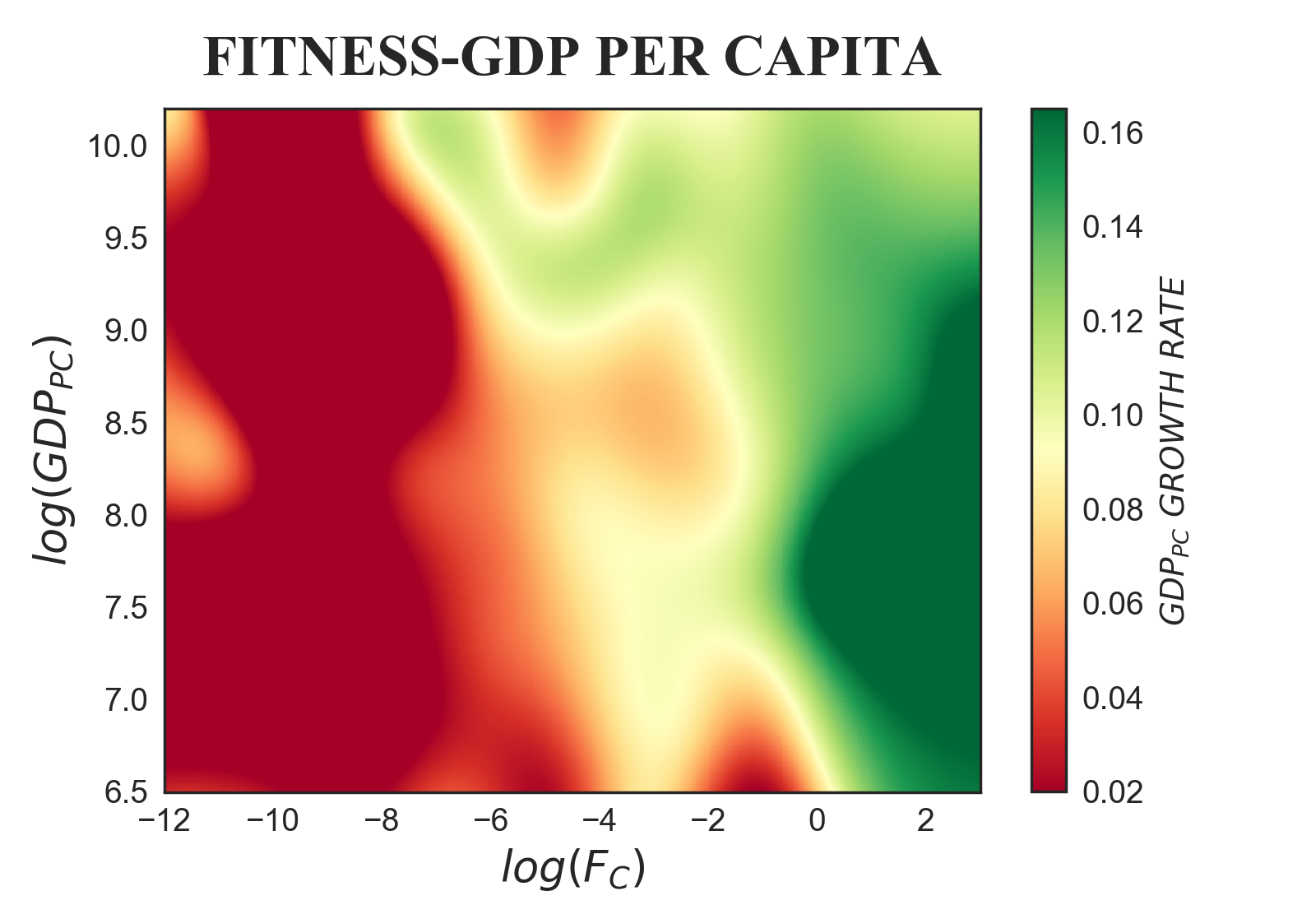}
\caption{\small The color-map represents the tridimensional relation between fitness, GDP per capita and subsequent GDP per capita growth rate, where a $\Delta t=5 $ years is considered. The variation of the growth rate is represented through color. The color-map is obtained through a non-parametric Nadaraya-Watson estimation. Countries with low fitness are not able to achieve subsequent high growth rates, irrespective of their initial GDP per capita level. For countries with intermediate initial fitness, higher starting GDP per capita results in mildly higher subsequent  growth rates. Finally, countries with high initial fitness are able to grow at very high rates, especially when their starting GDP per capita is low or intermediate. As has been  explored in detail [\citen{cristelli15, cristelli2017predictability, pugliese2017complex}], fitness, when put into relation with GDP per capita, is able to suggest future scheme of development not fully captured by solely monetary metrics.  \label{fig:gdp}}
 \end{figure}

Our empirical finding show that fitness is decisive in determining the growth profile of countries [\citen{pugliese2017complex,cristelli15,cristelli2017predictability}].
Looking at  the variability of color, in Figure \ref{fig:gdp} we can distinguish three areas. First, in the case of countries with low starting fitness ($-12<log(F_c ) \leqslant -7$, intense red), the GDP growth rate is completely led by fitness. When fitness is too low, independently of the initial income level, the take-off does not occur. A critical degree of complexity is necessary to start the catch-up [\citen{pugliese2017complex}]. 
Second, in the portion of the plot with countries which exhibit intermediate starting fitness ($-7<log(F_c)<0 $, orange and yellow), contrary to the predictions of exogenous growth theory, higher starting GDP per capita results in mildly higher growth rates.  Here, for example, we find oil exporters whose high income does not  reflect the high endowment of capabilities that would be necessary to sustain high economic growth.

Thirdly, for countries with high starting fitness ($log(F_c) \geqslant 0$, different shades of green), the convergence hypothesis seems to apply only when the initial fitness level is quite high . This brings to light a complementarity between the role of fitness and GDP per capita that cannot be ignored.  Countries with low initial GDP per capita and high fitness, have large capability portfolios which, after a critical threshold, enable them to catch-up and achieve very fast growth rates.  As has been pointed out in numerous recent contributions [\citen{cristelli15, cristelli2017predictability, pugliese2017complex}],  in this area of the fitness-GDP per capita plane we can find, among others, China and South Korea.

In Cristelli et al [\citen{cristelli15}] and in different other recent contributions [\citen{cristelli13, pugliese2017complex,cristelli2017predictability}], it has been explored in depth how fitness, when put into relation with GDP per capita, is a very powerful instrument in assessing the growth potential of countries that is not captured by the sole information contained in the GDP per capita.  Indeed fitness, being a structural measure that quantifies the capability endowment of countries, is able to profile them in a much more nuanced and coarse-grained way with respect to the aggregate measures of  growth inputs.


\subsubsection*{Fitness and Capital Intensity}

\begin{figure}[h]
\centering 
\includegraphics[width=.84\linewidth]{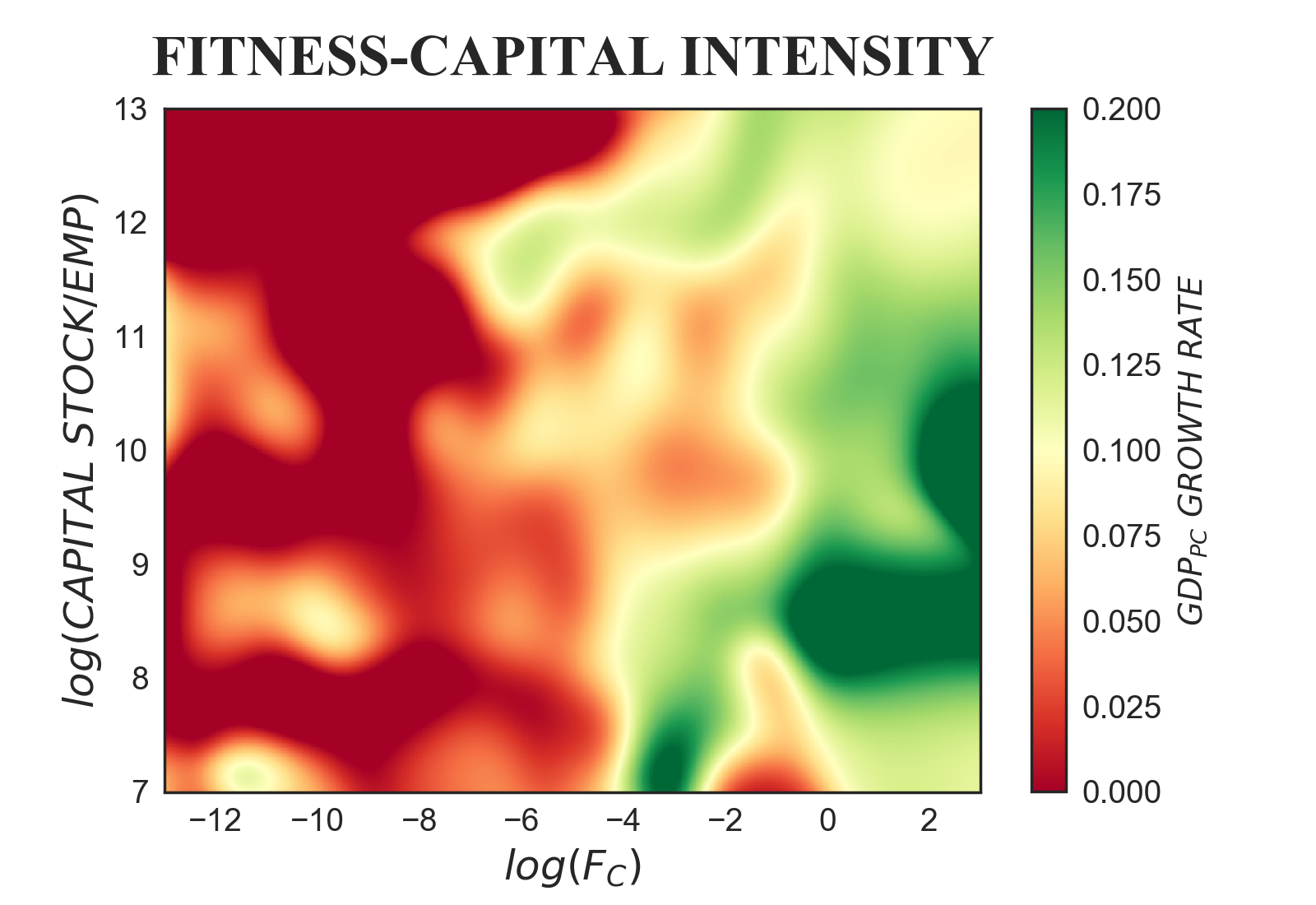}
\caption{\small
The color-map represents the tridimensional relation between fitness, capital intensity and subsequent GDP per capita growth rate, where a $\Delta t=5 $ years is considered. The variation of the growth rate is represented through color. The color-map is obtained through a non-parametric Nadaraya-Watson estimation. When the combined effect of fitness and capital intensity is taken into consideration, the latter looses explanatory power and the growth profile of countries is almost completely explained by their fitness level. Higher fitness  leads to higher growth rates, however, countries with high fitness and intermediate capital intensity are able to achieve the highest growth rates. \label{fig:kpc}}
 \end{figure}

Capital intensity is defined as the ratio between output and labor, \textit{i.e.} the number of workers employed in the economy. 
Under the neoclassical convergence hypothesis, in the long-run capital intensive  societies should tend to have higher standards of living, with higher GDP per capita levels and lower growth rates. By contrast, emerging economies should tend to be labour intensive  and show higher growth rates. Accordingly,  from the bottom to the top of our color-map we should observe that as capital intensity increases the color goes  from green (higher growth rate) to red (lower growth rate). 
However, in Figure \ref{fig:kpc}, when the combined effect of fitness and capital intensity  is taken into consideration the contour lines are mostly vertical, and the explanatory power of $K/EMP$ loses importance relative to fitness. Nevertheless, a partial reconciliation with the neoclassical models is found for very high starting fitness countries. The latter, similarly to what was examined in Figure \ref{fig:gdp}, countries with low/intermediate starting capital intensity are able to achieve the fastest growth rate. 

\subsubsection*{Fitness and Employment Rate}

The employment rate is defined as the employment-to-population ratio, where employment levels are include all the persons working within national boundaries. 

\begin{figure}[h]
\centering 
\includegraphics[width=.84\linewidth]{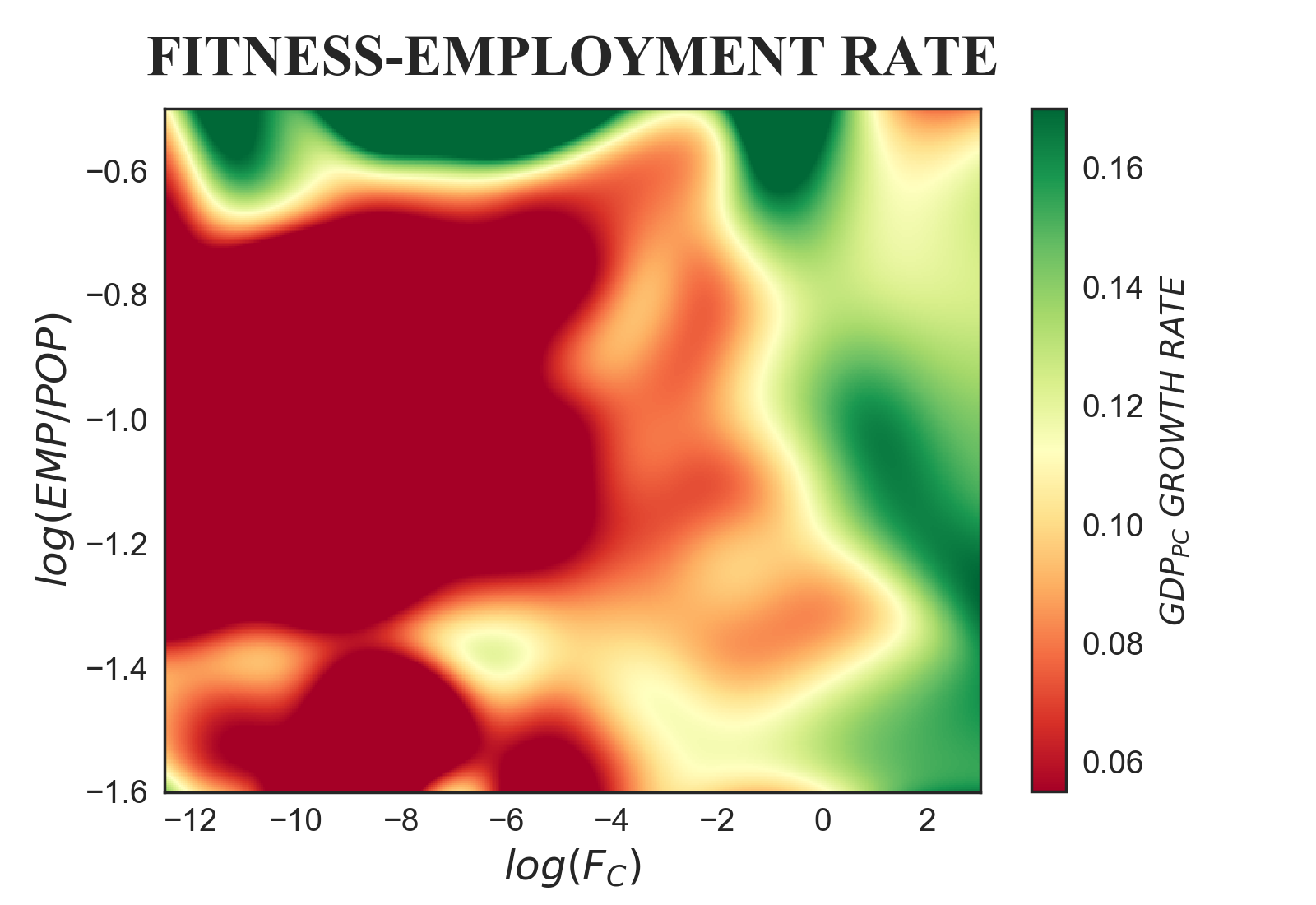}
\caption{\small The color-map represents the tridimensional relation between fitness, employment rate  and subsequent GDP per capita growth rate, where a $\Delta t=5 $ years is considered. The variation of the growth rate is represented through color. The color-map is obtained through a non-parametric Nadaraya-Watson estimation.
 Only the highest levels of employment rate, after a critical threshold of $EMP/POP\sim 55\%$, have a positive effect on economic growth. This is clearly visible from the horizontal variation of color from red to green in upper portion of the plot. For lower values of $EMP/POP$, fitness is the most important variables since the dominant variation of the color is vertical. The higher the fitnessm the  higher the GDP per capita growth rate.
\label{fig:emp}}
 \end{figure}

As mentioned in the Introduction, according to endogenous growth models, the employment rate should have a positive effect on economic growth. In particular,  Arrow's learning-by-doing model [\citen{arrow1962economic}] predicts that higher employment rates create positive externalities due to a larger size of the market.
In Figure \ref{fig:emp},  the movement of color hints at the presence of non-linear dependencies, and different behaviors of GDP per capita growth rate  can be identified. The positive effect of the employment rate on economic growth is detected only for the highest levels of employment rate, after a critical threshold of $EMP/POP\sim 55\%$, as can be seen from the horizontal variation of color from red to green in left-upper portion of the plot. For lower values of the employment rate, the dominant variation of the color is vertical: higher fitness leads to higher GDP per capita growth rates.

\subsubsection*{Fitness and Life Expectancy} 

Life expectancy is a proxy for healthcare quality. The effect of life expectancy on economic outcomes is positive and significant  according to the growth regressions by Barro [\citen{barro1991economic,barro2004economic}]: better health should lead to higher economic growth.

\begin{figure}[h]
\centering 
\includegraphics[width=.85\linewidth]{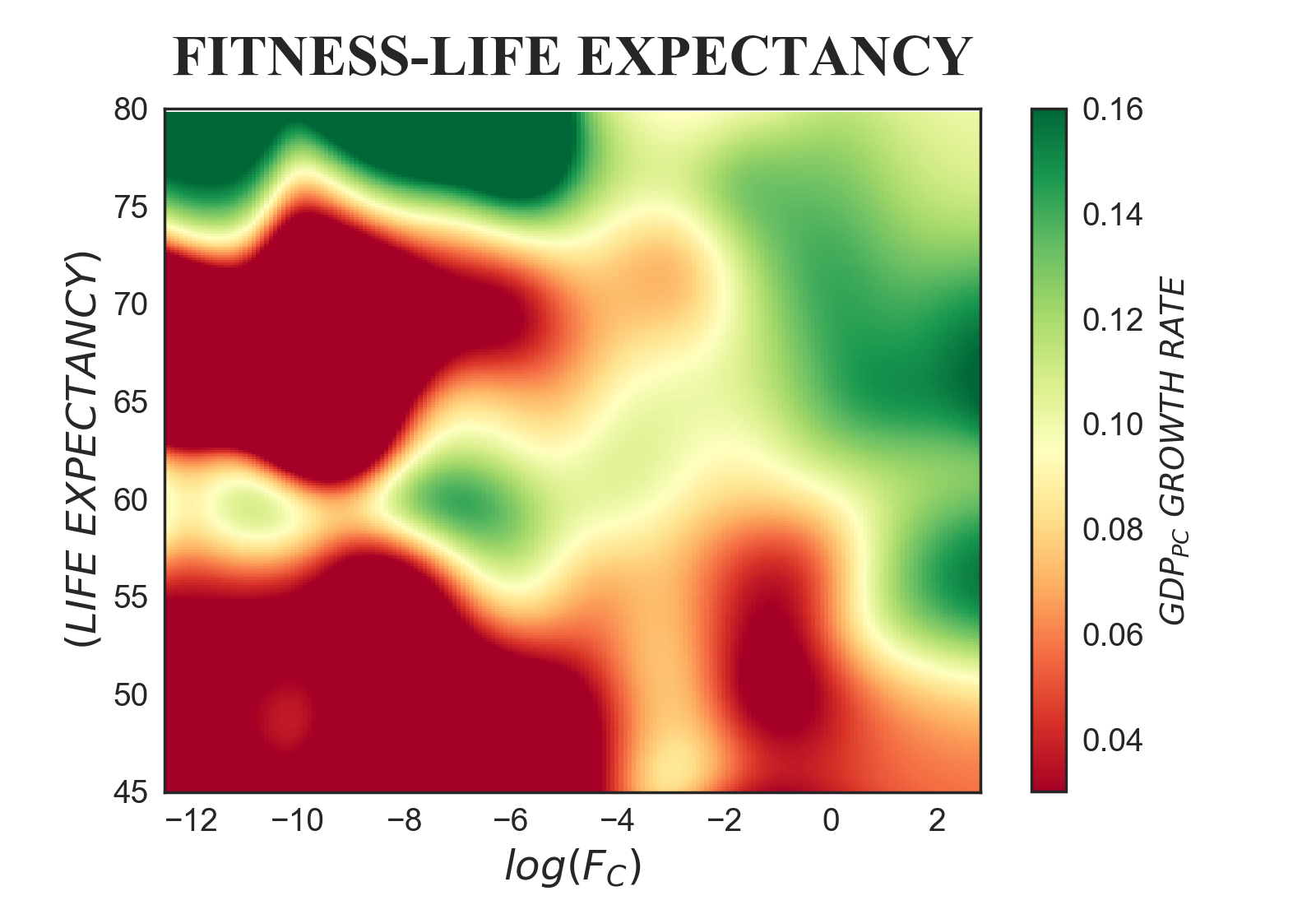}
\caption{\small The color-map represents the tridimensional relation between fitness, life expectancy and subsequent GDP per capita growth rate, where a $\Delta t=5 $ years is considered. The variation of the growth rate is represented through color. The color-map is obtained through a non-parametric Nadaraya-Watson estimation. Life expectancy values $ \gtrsim 73$ years  have a positive effect on growth rates. When  life expectancy $ <73$ years, fitness determines the color contour: the higher the fitness, the higher the growth rate. However, also high fitness countries show higher  growth rate when life expectancy $>60$ years. \label{fig:life_exp}}
 \end{figure}

In Figure \ref{fig:life_exp} such positive effect is only obtained when a critical threshold of \textit{circa} $73$ years  is exceeded. Similarly to Figure \ref{fig:emp}, also here a non-linear behavior is found, and fitness almost entirely dominates the dynamics.  Except for life expectancy $ \gtrsim 73$ years, low fitness countries grow slowly, whilst the contrary is true for high fitness countries. Nevertheless, for high fitness countries too growth rates are higher if life expectancy is approximately higher than 60 years. Such threshold falls for very high fitness countries: lower life expectancy is offset by increasing capabilities. 

However, when it comes to demographic variables, setting the direction of the arrow of causality may poses some problems. Life expectancy is listed among the growth determinants, but also the opposite relation is likely, it could be the case that high growth rates lift up life expectancy rather than the opposite.

 \subsubsection*{Fitness and Human Capital}
 In endogenous growth theories, human capital is an essential ingredient to achieve sustained growth rates. 
The accumulation of human capital, through formal training or learning-by-doing, creates positive externalities through increased productivity and technological innovation originating  from the process [\citen{arrow1962economic,lucas1988mechanics,aghion1992model}]. In Barro's growth regressions [\citen{barro1991economic, barro2004economic}] human capital is positively and significantly related to subsequent GDP per capita growth rates. 

\begin{figure}[h]
\centering 
\includegraphics[width=.85\linewidth]{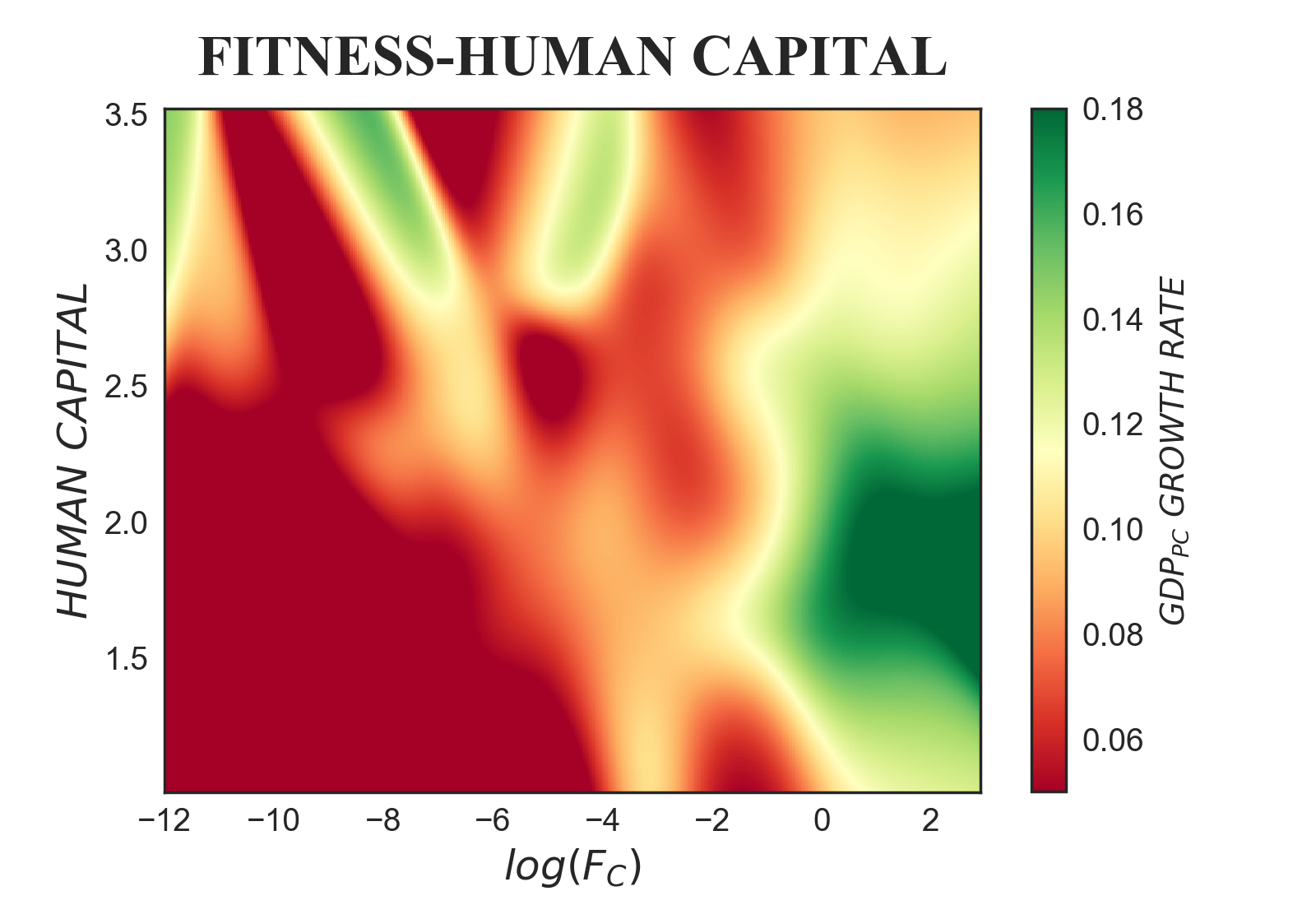}
\caption{\small The color-map represents the tridimensional relation between fitness, human capital and subsequent GDP per capita growth rate, where a $\Delta t=5 $ years is considered. The variation of the growth rate is represented through color. The color-map is obtained through a non-parametric Nadaraya-Watson estimation. Fitness and human capital appear positively correlated in some zones of the plots, and complementary in others. 
Low and intermediate fitness corresponds to low human capital. While, when  $log(F_c)>-4$ increasing initial fitness affects positively the GDP per capita growth rate, even for low initial human capital.   \label{fig:hc}}
 \end{figure}

Here human capital is proxied by a measure of educational attainment at each starting period. 
Figure \ref{fig:hc} puts into light a positive relation  between fitness and human capital (as confirmed by their significant Pearson correlation of $0.30$), principally for intermediate and high fitness values. In some areas of the figure fitness and human capital are entangled and reinforce each other, while in others high initial fitness can compensate for a lack of initial human capital. 


In fact, low and intermediate fitness countries are associated with
 low human capital ($<2$). By contrast, when $log(F_c)>-4$, increasing fitness  has a positive effect on economic growth even if human capital is low.

For $log(F_c)>-2$,  human capital takes off and 
the curve representing the fitness-human capital relation starts to assume an increasing shape. In this portion of the plot, we partially recover the expected effect of human capital on economic growth, fitness, though, still maintains a leading position. Notably, 
two main features of the relationships can be highlighted. First, as expected,  countries with very high human capital levels grow slowly, as can be seen  in the upper right corner of the figure. Second, in the bottom right corner are placed countries that, having very high fitness, are able to enter into sustained growth regimes even with low initial human capital levels. In such a way -as can be appreciated by the vertical passage from red to green, and then back to red- for high and intermediate fitness countries we observe a sort of inverted U-shaped relationship between human capital and the growth rate. Low initial human capital predicts low growth rate (lower red-yellow area), intermediate human capital predicts high growth rates (intense green area), and, finally, when human capital is high, the subsequent growth rates are low (upper orange area), but not as in the first phase. 

\subsubsection*{Fitness and Total Factor Productivity}
According to exogenous growth theories, total factor productivity (TFP) -or the Solow residual-  indirectly captures technological improvements. In the neoclassical production function  is identified as the growth of output that cannot be accounted for by the growth of the observed inputs, capital and labour. 
In this framework, an increase in total factor productivity should shift upwards the growth rate. 
Note that being estimated as a residual in the decomposition of the growth rate, all the measurement errors in capital and labour will be reflected in total factor productivity 

\begin{figure}[h]
\centering 
\includegraphics[width=.85\linewidth]{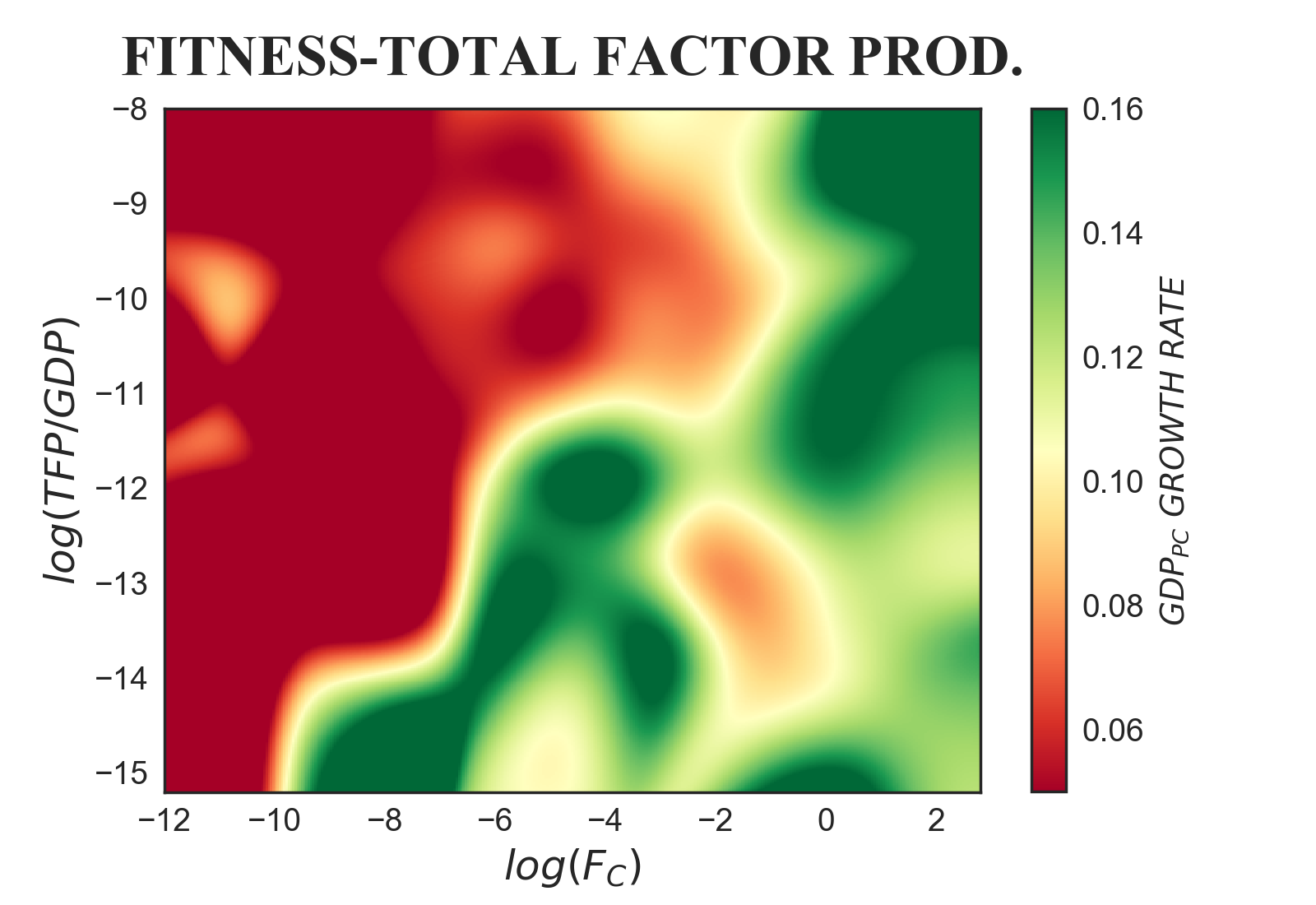}
\caption{\small The color-map represents the tridimensional relation between fitness, total factor productivity and subsequent GDP per capita growth rate, where a $\Delta t=5 $ years is considered. The variation of the growth rate is represented through color. The color-map is obtained through a non-parametric Nadaraya-Watson estimation. For $log(TFP/GDP) \gtrsim -11$,  as can be noted by the horizontal movement of color from red to green, fitness is the prevaling factor. Keeping $TFP$ constant, high initial fitness corresponds to greater subsequent  growth rates. For $log(TFP/GDP) < -11$ from the diagonal variation of color we can deduce that fitness and total factor productivity are complementary in affecting future growth rates. In this area, very low initial fitness brings low growth, independently from total factor productivity. For low and intermediate fitness countries, starting total factor productivity has a negative impact on economic growth. This could be due to total factor productivity 
measurement errors. Finally, high starting fitness leads to high growth rates, especially when total factor productivity is high. This highlights a complementarity of fitness and total factor productivity for highly competitive economies.   \label{fig:tfp}}
 \end{figure}

In the upper portion of Figure \ref{fig:tfp}, for $log(TFP/GDP) \gtrsim -11$,  as can be noted by the horizontal movement of color from red to green, the dynamic is dominated by fitness. Keeping $TFP$ constant, the higher the initial fitness, the greater the growth rate.  
Whereas, for $log(TFP/GDP)<-11$, as is apparent from the diagonal variation of color, even if fitness remains the prevailing factor, a complementarity between the effects of fitness and of total factor productivity slowly emerges. This area can be divided into three zones according to the corresponding fitness levels.
Firstly, for very low levels of initial fitness  ($-12<log(F_c)<-10$) we observe low growth rates, irrespective of the level of initial total factor productivity.   Secondly, for low and  intermediate levels of fitness  ($-10<log(F_c)<-2$), we observe a negative effect of initial total factor productivity on economic growth.  This might be attributed to measurement errors possibly due to an overestimation of the labour and capital factors. 
The source of such overestimation might be a misattribution of capabilities  to capital or labour. 
Thirdly, for $log(F_c)>-2$, growth rates are generally higher. However, the vertical transition of color from yellow to green shows that, as suggested by the theory, total factor productivity positively affects  the growth rate. Therefore, in this region of the plane, fitness and total factor productivity are mutually reinforcing.
 
In the non-parametric kernel regressions, which do not assume any aprioristic functional forms for the relationships, we have observed the combined effect of fitness, a proxy for the basket of capabilities of a country, and some key growth determinants. Our findings are unambiguous: fitness is an important, if not the most important, player in enhancing economic growth. Depending on the drivers and on the different phases that an economy is going through, when fitness is included in the analysis it can be either complementary to traditional drivers of growth or can completely overshadow them. The complementarity emerges in particular for high fitness countries, which have already a rich basket of capabilities.  This suggests that, once it becomes possible to account for capabilities, some other factors, otherwise relevant in determining future growth rates, lose importance. Not including fitness among explanatory variables in growth regressions may significantly diminish the overall explanatory power of the 
model. Additionally it may cause an omitted variable bias and thus produce misleading results; the significance of some regression coefficients may in fact be spurious and be lost when also the effect of fitness is considered.

Different endowment of underlying  capabilities explain differences in economic performances,  greater complexity enable countries to use factors more efficiently, be more competitive and catch-up more rapidly [\citen{sutton2002rich, sutton2012competing, sutton2016capabilities, cristelli2017predictability,pugliese2017complex}].


\subsection{Econometric Analysis}

Now we perform a parametric linear regression using as explanatory variables GDP per capita $\log(\mathrm{GDPpc})$, Capital Intensity $\log(\mathrm{K/EMP})$, number of employees $\log(\mathrm{EMP})$, the inverse of Life Expectancy $\log(\mathrm{1/LifeExp})$, Secondary Schooling $\log(\mathrm{School})$, and Total Factor Productivity $\log(\mathrm{TFP/GDP})$. The model is therefore
\begin{multline}
\log(\mathrm{GDPpc}) \sim \mathrm{const.}+\beta_1 \log(\mathrm{K/EMP})+\beta_2 \log(\mathrm{EMP})+\beta_3 \log(\mathrm{TFP/GDP})+ \\
\beta_4 \log(\mathrm{1/LifeExp})+\beta_5 \log(\mathrm{School})+\beta_5 \mathrm{Fitness Rank}
\end{multline}

Various drivers of economic growth figure as right-hand-side explanatory variables affecting the growth rate of GDP per capita, on the left-hand side of the equation. The idea is to add fitness to the drivers and observe whether and to which extent the results change. The results of our analysis are shown in Table \ref{tab:regr}. The Fitness Rank is highly significant. The effect of Life Expectancy is positive and significant, while the coefficient of the GDP per capita is negative. Both results are in agreement with the literature; in particular, the second one is coherent with the convergence scheme reported by Barro [\citen{barro1991convergence}] and Mankiw et. al [\citen{mankiw1992contribution}]. The other variables display an unexpected behaviour, probably attributable to measurament errors (Schooling, TFP), low input substitution (Schooling) or multicollinearity (number of employees, Schooling, Capital Intensity).
We are planning to investigate further this parametric approach, possibly taking into account also fixed effects in the regression model.


\begin{table}
\centering
\begin{tabular}{l|l|l}
Variable & Coefficient & Standard Error \\\hline
Constant & -1.44 & 0.35\\
$\log(\mathrm{GDPpc})$ & -0.084 & 0.028 \\
$\log(\mathrm{K/EMP})$ & -0.025 & 0.016 \\
$\log(\mathrm{EMP})$ & -0.051 & 0.027 \\
$\log(\mathrm{TFP/GDP})$ & -0.048 & 0.028 \\
$\log(\mathrm{1/LifeExp})$ & -0.49 & 0.11 \\
$\log(\mathrm{School})$ & -0.042 & 0.020 \\
Fitness Rank & 0.255 & 0.048
\end{tabular}
\begin{tabular}{l|l|l}
Variable & Coefficient & Standard Error \\\hline
Constant & -2.36 & 0.38\\
$\log(\mathrm{GDPpc})$ & -0.069 & 0.032 \\
$\log(\mathrm{K/EMP})$ & -0.025 & 0.019 \\
$\log(\mathrm{EMP})$ & -0.020 & 0.029 \\
$\log(\mathrm{TFP/GDP})$ & -0.035 & 0.031 \\
$\log(\mathrm{1/LifeExp})$ & -0.73 & 0.12 \\
$\log(\mathrm{School})$ & -0.036 & 0.020 
\end{tabular}
\caption{\label{tab:regr}Parametric regression results, including or not the Fitness Rank as an explanatory variable (left and right table, respectively).}
\end{table}

\section{Conclusion}\label{sec:concl}
This paper has examined how complex network analysis can be a powerful tool to understand economic growth. The traditional literature on economic growth does not fully consider the role of the underlying capabilities, which are however crucial for the adoption of advanced technologies and for the introduction of innovations. The Fitness metric used in the paper, which captures the global export specialization profile of countries, is shown to be a significant variable for predicting growth. Economic models of growth that overlook the role of complexity therefore run the risk of being misspecified.


%

\vspace{6pt} 




\section*{Acknowledgements and funding}{We would like to thank Luciano Pietronero for comments and discussions. A.Z. acknowledges funding from the 'PNR Progetto di Interesse' CRISIS LAB. 

The founding sponsors had no role in the design of the study; in the collection, analyses, or interpretation of data; in the writing of the manuscript, and in the decision to publish the results. The authors declare no conflict of interest.}

\end{document}